\documentclass[a4paper,twocolumn]{article}
\usepackage{xcolor}
\usepackage{graphicx}
\usepackage[super,comma,sort&compress]{natbib}
\usepackage{epstopdf}
\usepackage[margin=2cm]{geometry}
\usepackage{amsmath}
\usepackage{mhchem}
\usepackage{caption}
\begin{document}

\title{Resilience of small PAHs in interstellar clouds:\\Efficient stabilization of cyanonaphthalene by fast radiative cooling}


\author{Mark H. Stockett$^{1*}$, James N. Bull$^2$, Henrik Cederquist$^1$, Suvasthika Indrajith$^1$,\\
MingChao Ji$^1$, Jos{\'e} E. Navarro Navarrete$^1$, Henning T. Schmidt$^1$, Henning Zettergren$^1$\\ \& Boxing Zhu$^1$}
\date{$^1$Department of Physics, Stockholm University, Stockholm, Sweden\\
$^2$School of Chemistry, University of East Anglia, Norwich, United Kingdom\\
$^*$Mark.Stockett@fysik.su.se}

%
\maketitle
\begin{abstract}
After decades of speculation and searching, astronomers have recently identified specific Polycyclic Aromatic Hydrocarbons (PAHs) in space. Remarkably, the observed abundance of cyanonaphthalene (CNN, \ce{C10H7CN}) in the Taurus Molecular Cloud (TMC-1) is six orders of magnitude higher than expected from astrophysical modeling. Here, we report absolute unimolecular dissociation and radiative cooling rate coefficients of the 1-CNN isomer in its cationic form. These results are based on measurements of the time-dependent neutral product emission rate and Kinetic Energy Release distributions produced from an ensemble of internally excited 1-CNN$^+$ studied in an environment similar to that in interstellar clouds. We find that Recurrent Fluorescence -- radiative relaxation via thermally populated electronic excited states -- efficiently stabilizes 1-CNN$^+$, owing to a large enhancement of the electronic transition probability by vibronic coupling. Our results help explain the anomalous abundance of CNN in TMC-1 and challenge the widely accepted picture of rapid destruction of small PAHs in space.
\end{abstract}


\section{Introduction}

Polycyclic Aromatic Hydrocarbons (PAHs) have long been thought to be ubiquitous in the Interstellar Medium, as evidenced by the infrared (IR) emission bands observed by astronomers at wavelengths coincident with their vibrational transition energies \cite{Li2020}. These partially resolved bands, however, are common to PAHs as a class of molecules and cannot be used to identify specific species. To date, it has been generally held that PAHs in the interstellar medium must be fairly large, containing more than 50 carbon atoms\cite{Li2020} to be resilient against fragmentation after collisions or photon absorption. Here, we present experimental results demonstrating that a small PAH cation is stabilized much more rapidly than previously assumed, and that this occurs efficiently through recurrent florescence.   

\begin{figure}
\centering
\includegraphics[width=0.6\columnwidth]{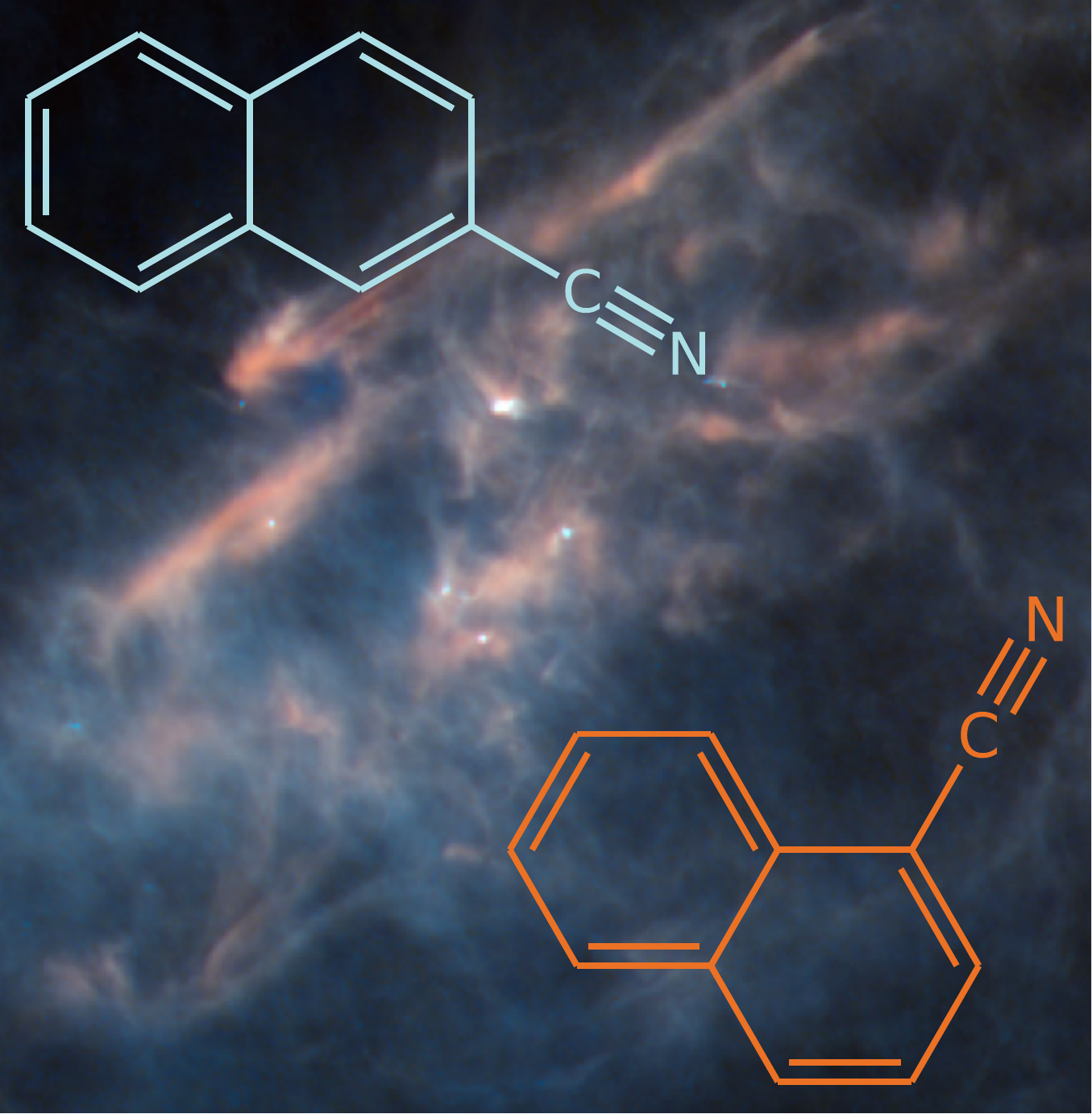}
\caption{\textbf{Found in space.} Molecular structures of 1-cyanonaphthalene (1-CNN, orange) and 2-CNN (blue), \ce{C10H7CN}. Background: Image of TMC-1 from ESA.}
\label{fig_struct}
\end{figure}

Recently, McGuire \textit{et al.} \cite{McGuire2021} analyzed radio telescope observations of the dark molecular cloud TMC-1 to identify two isomers of cyanonaphthalene (\ce{C10H7CN}, Fig.~\ref{fig_struct}) -- in which a nitrile/cyano (-CN) group replaces one of the hydrogen atoms of naphthalene. This is the first definitive assignment of a specific PAH molecule in space. As important as this assignment is for finally confirming the presence of PAH molecules in space, it is equally remarkable that the observed CNN abundances, as well of that of the smaller but structurlly similar benzonitrile, are orders of magnitude higher than predicted by current astrochemical modeling \cite{McGuire2018,McGuire2021}. While laboratory studies have shown that nitrogen bearing aromiatic molecules \cite{Rap2022}, as well as pure hydrocarbons like naphthalene\cite{Parker2012} and indene\cite{Burkhardt2021,Doddipatla2021}, the model of McGuire \textit{et al.} indicates rapid depletion of CNN from TMC-1 through interactions with ions and limits the abundance of its main precurser, naphthalene, which has also been shown to be an keystone in the formation of larger PAHs\cite{Lemmens2020}. McGuire \textit{et al.} argue that radiative cooling of small PAH cations ($<20$ atoms) is too slow to stabilize the molecules following ionizing interactions \cite{Rapacioli2006,Montillaud2013}.  However, these arguments are based on the assumption that only radiative stabilization \textit{via} emission of IR photons from transitions between vibrational levels contribute to the cooling. Here, we use a cryogenic ion beam storage ring provding 'moleular cloud in a box' conditions to show that cooling is much faster, due to emission of optical photons from thermally populated electronically excited states, \textit{i.e.} Recurrent Fluorescence (RF). Thus, some of the destruction channels included in the modelling by McGuire \textit{et al.} \cite{McGuire2021} are avoided, leading to a significantly higher survival probability of 1-CNN$^+$ in interstellar clouds.


Investigations of Recurrent Fluorescence (RF) are an emerging theme of research within laboratory astrophysics. In a typical experiment, internally hot ions are produced in a plasma ion source or through laser excitation followed by internal conversion to give the vibrationally-excited electronic ground state \cite{Martin2013,Saito2020}. These hot ions may, by inverse internal conversion, spontaneously \cite{Nitzan1979} populate electronically excited states which may, in turn, relax by emitting optical photons. While a few outstanding experiments have succeeded in direct detection of RF photons \cite{Ebara2016,Saito2020}, most reports of RF have been inferred indirectly through the quenching of dissociation or other destruction channels on timescales that are too rapid to be explained by sequential emission of IR photons \cite{Martin2013,Ito2014,Chandrasekaran2014}. 

The dominant dissociation channel of 1-CNN$^+$ observed under our experimental conditions is \cite{West2019}:

\begin{equation}
\ce{C10H7CN+ -> C10H6+ + HCN}+\epsilon,
\label{eq_hcn}
\end{equation}
where $\epsilon$ is the kinetic energy released in the reaction. Time-dependent dissociation rates $\Gamma(t)$ and Kinetic Energy Release (KER) distributions $P(\epsilon)$ were measured for ensembles of 1-CNN cations produced with an initially broad internal energy distribution, $g(E,t=0)$, leaving the ion source at time $t=0$. The ensemble was stored in the cryogenic electrostatic ion-beam storage ring DESIREE \cite{Thomas2011,Schmidt2013} (see Sec.~\ref{sec_exp}). From analysis of the KER distributions, we determine the unimolecular dissociation rate coefficient $k_{diss}(E)$ as a function of the internal excitation energy $E$. We reproduce the measured absolute dissociation rate $\Gamma(t)$ with master equation simulations (Sec.~\ref{sec_model}) of the time evolution of the energy distribution $g(E,t)$ and calculated vibrational and electronic (RF) cooling rates, but only when we include Herzberg-Teller vibronic coupling in the RF rate coefficient. The implications of these results for the observed high abundance of 1-CNN in TMC-1 are discussed in Sec.~\ref{sec_disc}.

\section{Results}


\begin{figure}
\includegraphics[width=\columnwidth]{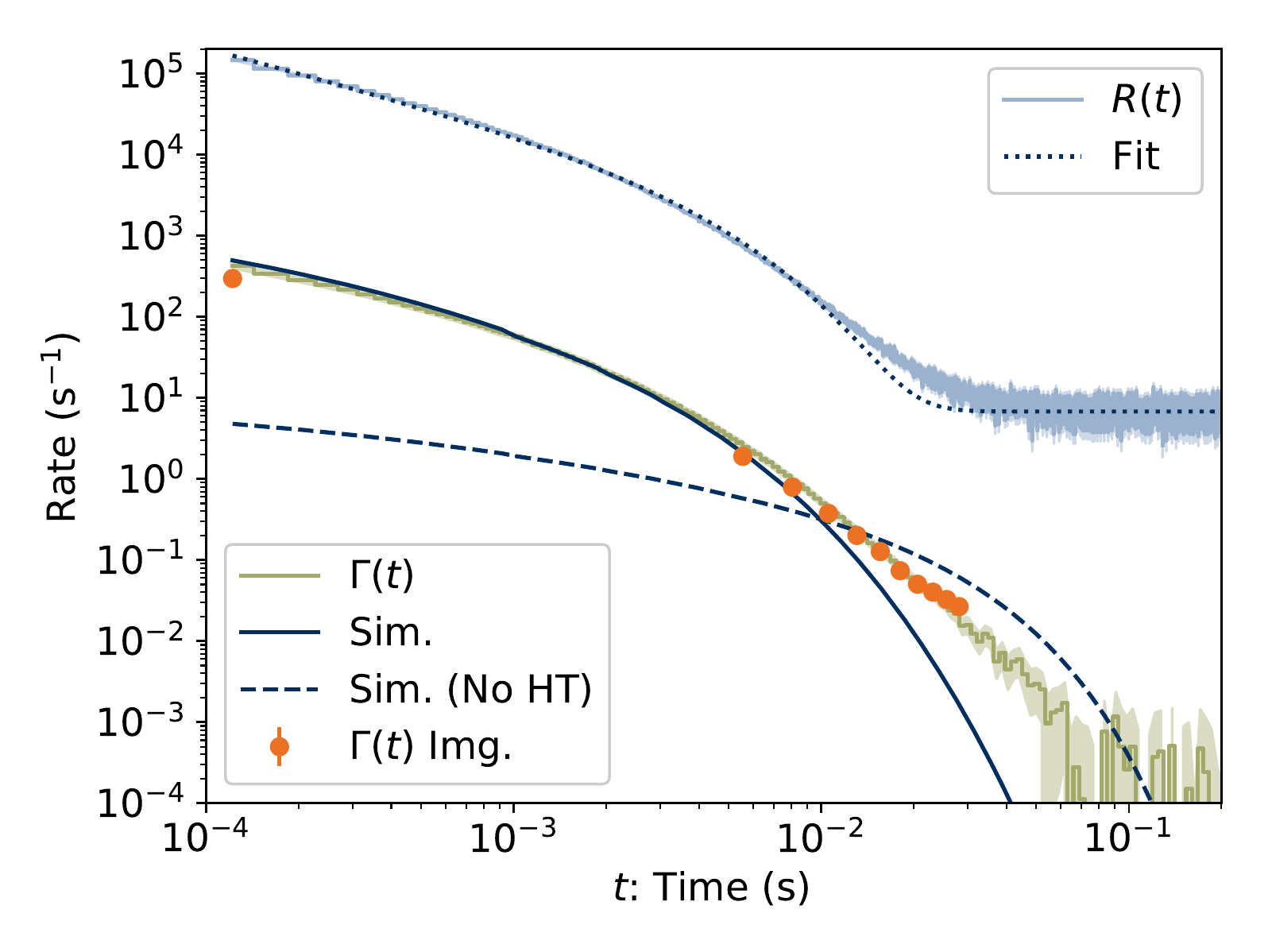}
\caption{\textbf{Time-dependent dissociation rate of ensembles of internally hot 1-CNN$^+$.} The storage time $t$ is relative to formation of the ions in the ion source. $R(t)$ is the rate of neutral product detection recorded continuously during storage of the ion beam. The dotted curve `Fit' is a fit of Eq.~\ref{eq_analfit} to $R(t)$. $\Gamma(t)$ is the absolute, ensemble-averaged dissociation rate. `$\Gamma(t)$ Img.' is the rate extracted from KER distributions measured at specific times. The error bands and bars are the standard deviations. The solid curve `Sim.' is the result of our master equation simulation including Herzberg-Teller coupling, which gives $f=0.011$, and the best-fitting initial temperature of 1860~K. The dashed curve `Sim. (No HT)' is a simulation without Herzberg-Teller coupling ($f=10^{-4}$) with a best-fitting initial temperature of 1310~K.}
\label{fig_gamma}
\end{figure}

The measured dissociation rate for ensembles of internally hot 1-CNN cations stored in DESIREE is shown in Fig.~\ref{fig_gamma}. During the measurement, the yield of neutral fragments (HCN molecules according to Eq.~\ref{eq_hcn}) leaving the storage ring is recorded continuously as a function of time $t$ after the ions left the source at $t=0$. The measured count rate, $R(t)$, is averaged over a large number of injection and storage cycles. During the first 10$^{-3}$ seconds after formation, the dissociation rate follows a power law $R(t)\propto t^{-1}$, as a consequence of the broad distribution $g(E,t)$ of internal energies, $E$, and rapid variation with $E$ of the dissociation rate coefficient $k_{diss}(E)$ \cite{Hansen2001}. After a critical time $k_c^{-1}$, the dissociation rate is quenched by competition with radiative cooling \cite{Andersen2001}, giving a dissociation rate with the approximate time dependence:

\begin{equation}
R(t)=r_0t^{-1}e^{-k_ct}.
\label{eq_analfit}
\end{equation}
The dotted curve labeled `Fit' in Fig.~\ref{fig_gamma} is a fit of this equation with a constant background term to the measured data $R(t)$. The critical rate coefficient, determined to be $k_c=300(20)$~s$^{-1}$, is an indication of efficient radiative stabilization of 1-CNN$^+$ \cite{Martin2013}. However, the value of $k_c$ alone does not provide any information on the cooling mechanism or the internal energy at which cooling and dissociation are competitive. For this, absolute dissociation rate coefficients need to be determined.  

The absolute per-particle dissociation rate is the ensemble average of the dissociation rate coefficient:

\begin{equation}
\Gamma(t)=\int g(E,t)k_{diss}(E)dE/\int g(E,t)dE.
\end{equation}
and is related to the measured count rate $R(t)$ by \cite{Schmidt2013,Stockett2020b}

\begin{equation}
\Gamma(t) = \frac{C}{\eta_{det} L_{SS}N(t)}R(t),
\label{eq_rgamma}
\end{equation}
where $\eta_{det}=0.34(3)$ (see Supplemental Information) is the efficiency for detection of HCN, $C=8.7$~m is the circumference of the storage ring, $L_{SS}=0.95$~m is the length of the stored beam viewed by the detector, and $N(t)$ is the average number of stored ions remaining in the ring at time $t$. The latter is determined from the decay rate $R(t)$, measured during ion storage, and the terminal ion beam current, measured at the end of each injection-storage cycle (see Supplemental Information). The absolute dissociation rate $\Gamma(t)$ for 1-CNN$^+$ is given in Fig.~\ref{fig_gamma}, with the background count rate due to collisions with residual gas (0.23(1)~s$^{-1}$) and detector dark noise (6.78(2)~s$^{-1}$) subtracted from the experimental data. 

The unitless factor $r_0C/\eta_{det} L_{SS}N(t)$ found when inserting Eq.~\ref{eq_analfit} into Eq.~\ref{eq_rgamma}, which is on the order of $10^{-1}$, gives the fraction of ions with non-negligible probability of decaying at time $t$ \textit{i.e.} those with internal energies $E$ where the inverse dissociation rate coefficient $k_{diss}(E)^{-1}\approx t$ \cite{Hansen2020}.


\begin{figure}
\includegraphics[width=\columnwidth]{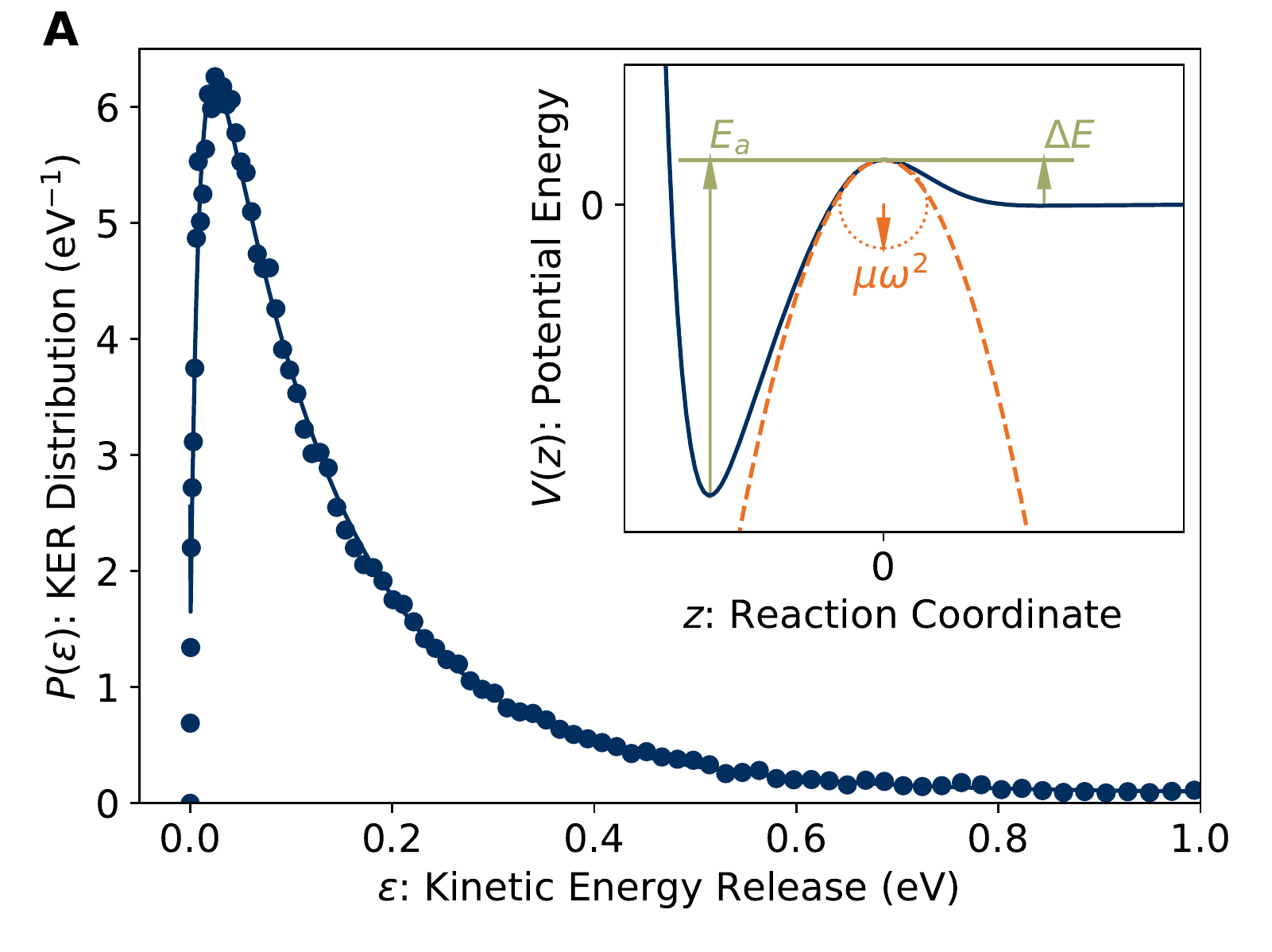}
\includegraphics[width=\columnwidth]{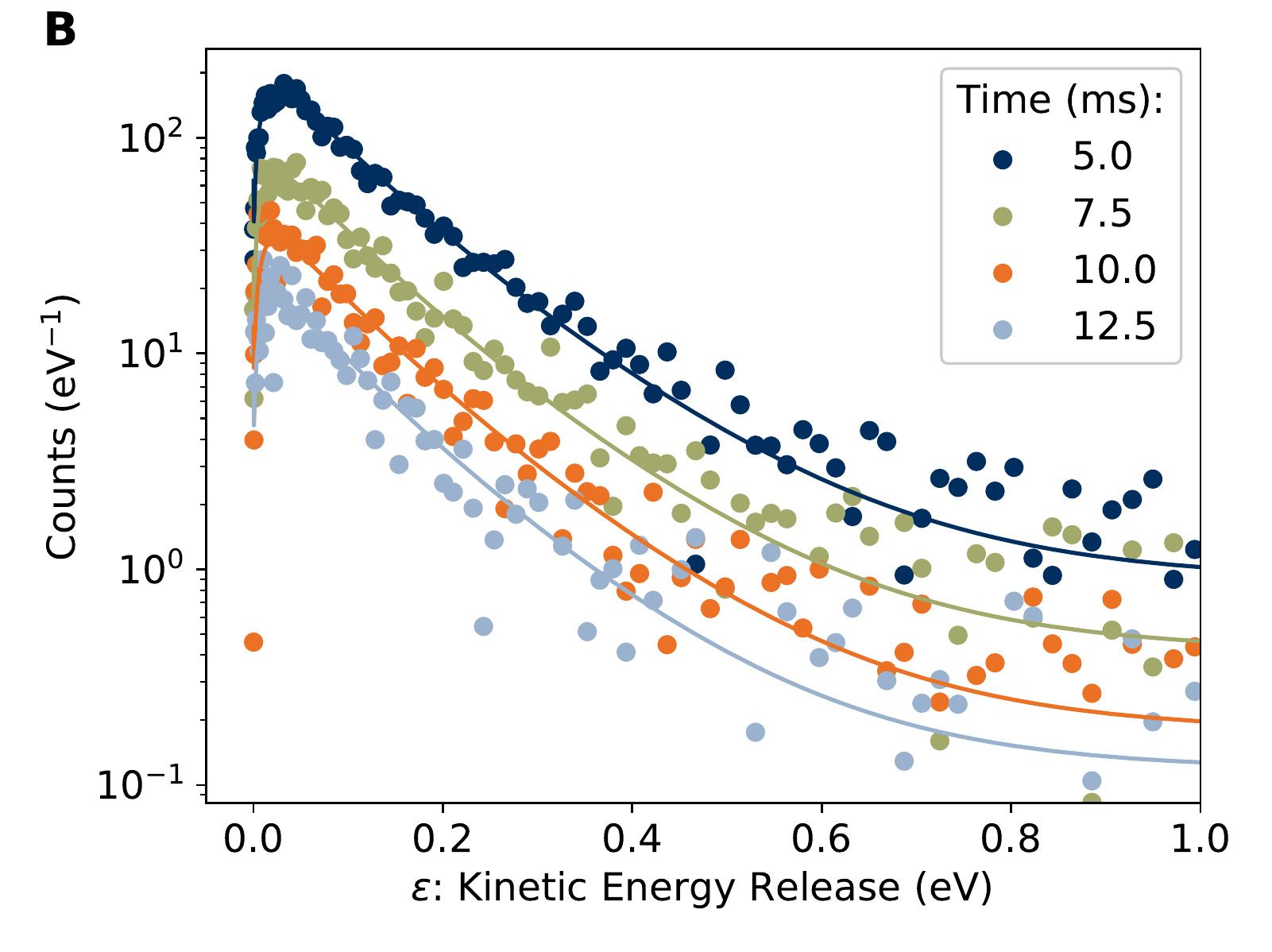}

\caption{\textbf{Kinetic Energy Release (KER) distributions for HCN-loss from 1-CNN$^+$.} \textbf{A} Normalized KER distribution recorded 120~$\mu$s after ion formation. The solid line is a fit of Eqn.~\ref{eq_kerd}. The inset illustrates the model parameters. \textbf{B} KER distributions recorded 5.0--12.5~ms after ion formation. Solid lines are the results of a simultaneous fit to the KER distributions from 120~$\mu$s to 20~ms. Time refers to the beginning of the $\Delta t=1.1$~ms camera exposure. Note the logarithmic vertical scale.}
\label{fig_ker}
\end{figure}

In order to connect the dissociation rate $\Gamma(t)$ to the internal excitation energy $E$, we analyze   the KER distributions to determine the unimolecular dissociation rate coefficient in Arrhenius form:

\begin{equation}
k_{diss}(E(T^{\ddag}))=A^{diss}e^{-E_a/k_BT^{\ddag}},
\label{eq_kd}
\end{equation}
where $E_{a}$ and $A^{diss}$ are the activation energy and pre-exponential factor of the decay channel, respectively, and $T^{\ddag}$ is the temperature of the transitory ion-molecule complex. The latter is related to the microcanonical excitation energy $E$ through the caloric curve (see Supplementary Information). In transition state theory, the pre-exponential factor is given by

\begin{equation}
A^{diss}=\frac{k_BT^{\ddag}}{h}e^{1+\Delta S^{\ddag}/N_Ak_B}
\end{equation}
where $\Delta S^{\ddag}$ is the activation entropy and $N_A$ is Avogadro's number \cite{Leyh1999}. 

The KER distribution measured for the ensemble of ions at $t=120$~$\mu$s is shown in Fig.~\ref{fig_ker}A. This KER distribution is well-reproduced by the model elaborated by Hansen \cite{Hansen2018}, which considers a transition state with activation energy $E_a$, reverse barrier height $\Delta E$, and a potential near the saddle point $V(z)\approx\Delta E-\frac{1}{2}\mu\omega^2z^2$, where $z$ is the reaction coordinate measured from the top of the barrier, $\mu$ is the reduced mass of the dissociation products (Eq.~\ref{eq_hcn}), and $\mu\omega^2$ is the radius of curvature of the potential near the saddle point. The model parameters are illustrated schematically in the inset to Fig.~\ref{fig_ker}A. The KER distribution takes the form \cite{Hansen2018}:

\begin{multline}
P(\epsilon) \propto \frac{e^{\beta'}}{e^{\beta'}+1}e^{-(\epsilon-\Delta E)/k_BT^{\ddag}}, \\ \mathrm{where}\ \beta'=4\pi\frac{\Delta E}{\hbar\omega}\left(\sqrt{\frac{\epsilon}{\Delta E}}-1\right).
\label{eq_kerd}
\end{multline}
The solid line in Fig.~\ref{fig_ker}A is a fit of Eq.~\ref{eq_kerd}, giving a temperature $T^{\ddag}=1610(20)$~K. The reverse barrier $\Delta E=6.2(5)$~meV, also obtained from the fit, is comparable to the centrifugal barrier for 1-CNN assuming the rotational temperature is of the same order as $T^{\ddag}$ \cite{Hansen2018}. Given such a `flat' transition state \cite{Leyh1999}, we invoke the simplifying assumption $\Delta S^{\ddag}\approx 0$, giving:

\begin{equation}
k_{diss}(E(T^{\ddag}))\approx\frac{k_BT^{\ddag}}{h}e^{-E_a/k_BT^{\ddag}},
\label{eq_arrhenius} 
\end{equation}
where the error introduced in $k_{diss}$ due to this approximation much smaller than the error due to the uncertainty in $E_a$. We next simultaneously fit the KER distributions recorded up to $t=20.0$~ms, subject to the constraint:

\begin{equation}
\int P(\epsilon)d\epsilon=R(t)\Delta t=r_0k_{diss}(E(T^{\ddag}))\Delta t 
\label{eq_keff}
\end{equation}
where $\Delta t$ is the camera exposure time, and $k_{diss}(E(T^{\ddag}))$ is given by Eq.~\ref{eq_arrhenius}. This fit includes contributions from ions dissociating at different points along the straight section of the storage ring seen by the detector. Results for $t=5.0$, 7.5, 10.0, and 12.5~ms are shown in Fig~\ref{fig_ker}B. The fitted temperatures for these $P(\epsilon)$ distributions range from 1320(10)--1220(10)~K. From the full data set, including $t=120$~$\mu$s, we obtain $E_a=3.16(4)$~eV. Our results are close to those of So \& Dunbar \cite{So1988} for benzonitrile (which is structurally similar to CNN), who determined $E_a=3.015$~eV, and of West \textit{et al.} \cite{West2019} who estimated an activation energy ``from 2.5 to 3~eV'' for 1- and 2-CNN. 

The factor $r_0$ enters Eq.~\ref{eq_keff} as the rate coefficient pertains to the fraction of ions with vibrational temperatures $T^{\ddag}$ at the observation time $t$, \textit{i.e.} $k_{diss}(E(T^{\ddag}))=t^{-1}$, which is equivalent to the definition of $r_0$ above. The values of $\Gamma(t)$ extracted from the imaging data (Eq.~\ref{eq_keff}) are shown in Fig.~\ref{fig_gamma}, labeled `$\Gamma(t)$ Img.', and agree well with those from the decay rate measurement (Eq.~\ref{eq_rgamma}). 

\begin{figure}
\includegraphics[width=\columnwidth]{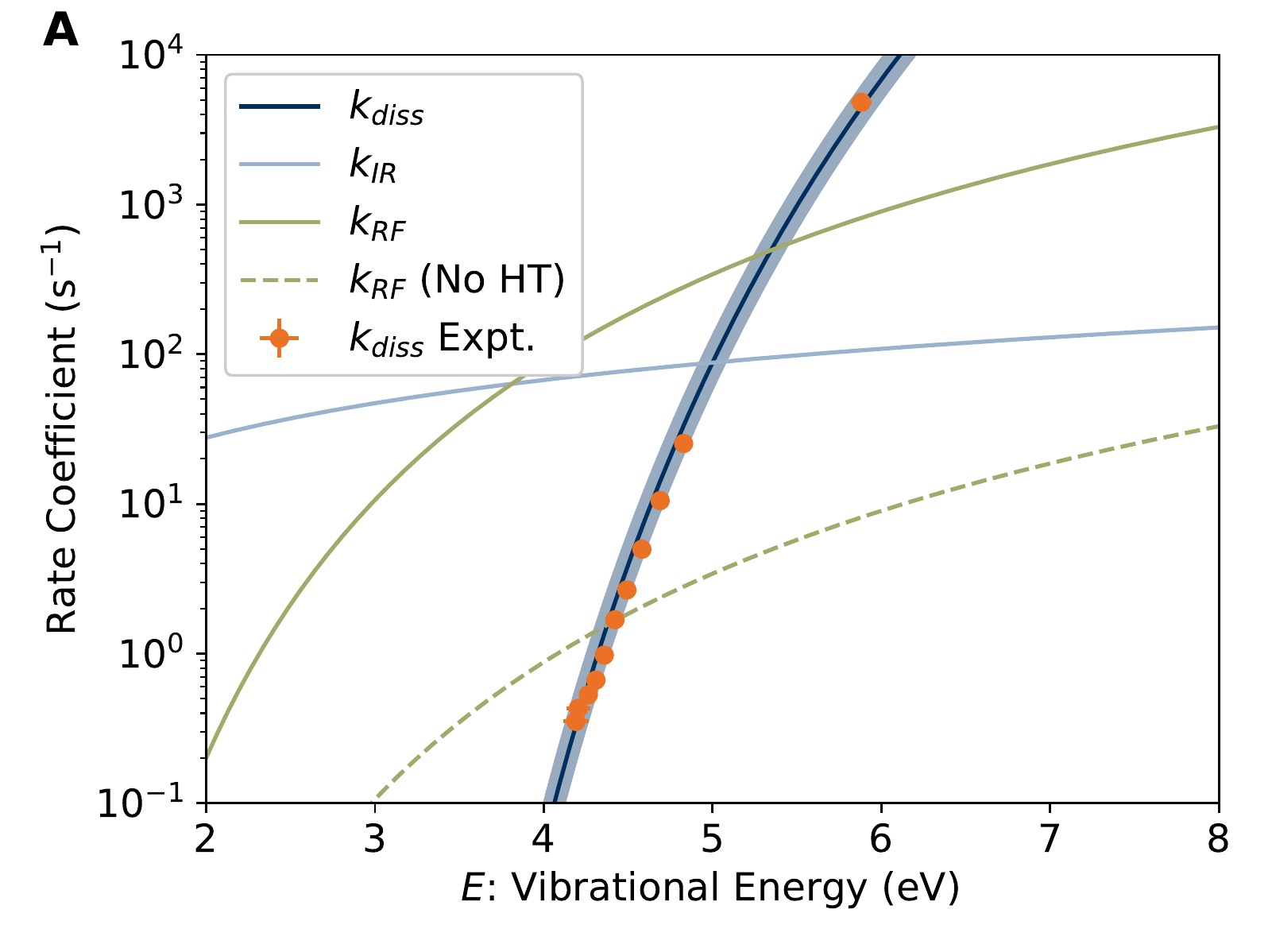}
\includegraphics[width=\columnwidth]{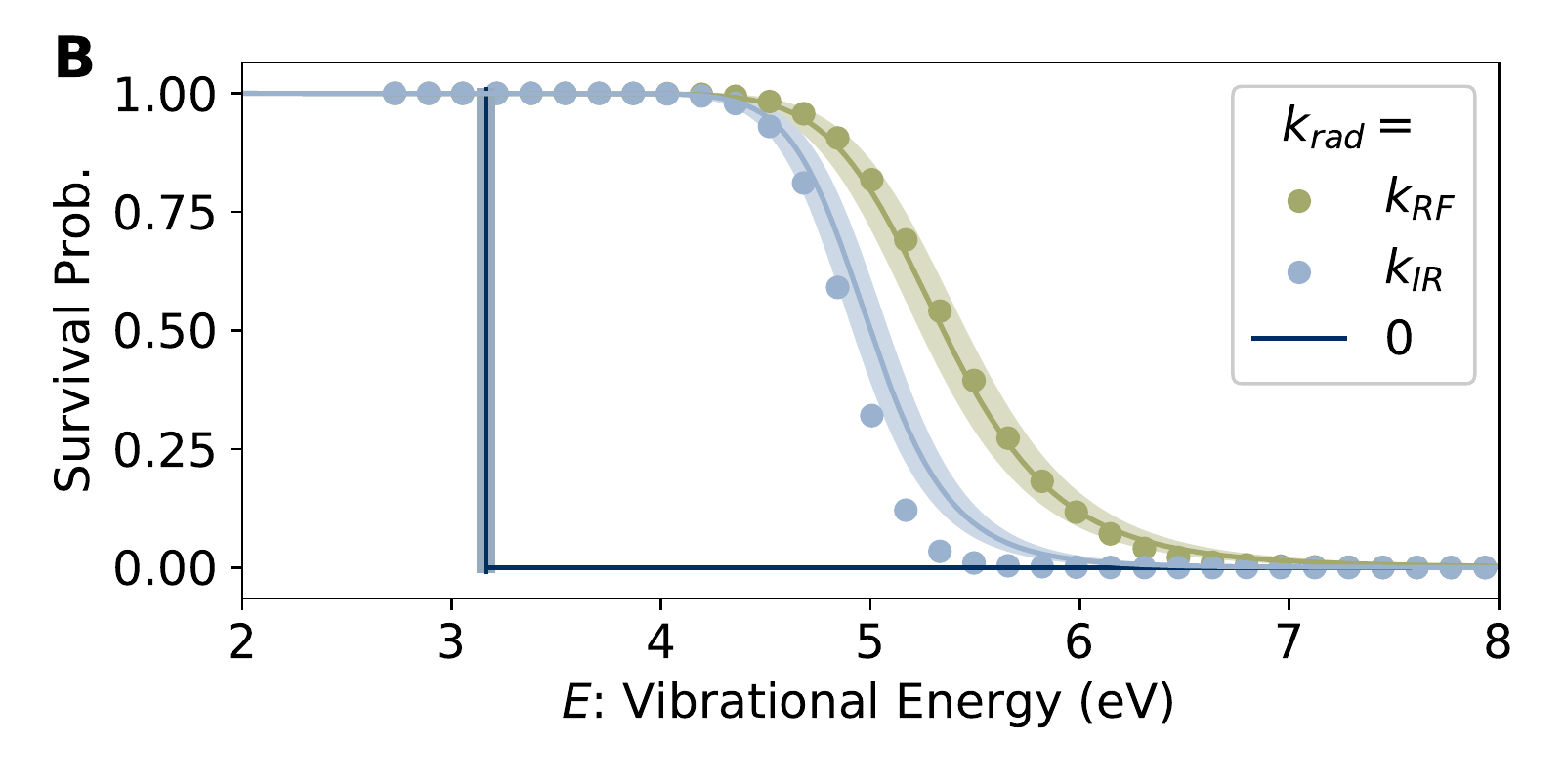}
\caption{\textbf{Rate coefficients and survival probabilities.} \textbf{A} Dissociation rate coefficient $k_{diss}$(E), fit to the experimental values derived from KER distributions, radiative cooling rate coefficients $k_{IR}(E)$ and $k_{RF}(E)$ using computed vibrational and electronic transition energies. The error bars are standard deviations of the fit parameter estimates. \textbf{B} Survival probability for 1-CNN$^+$ from master equation simulations (symbols) and the approximate expression $k_{rad}/(k_{diss}+k_{rad})$ (solid lines). Error bands in both panels reflect the uncertainty due to the standard deviation in the estimate of $E_a$.}
\label{fig_rates}
\end{figure}

Our measured dissociation rate coefficients from the imaging experiments are plotted in Fig.~\ref{fig_rates}, along with the dissociation rate coefficient in microcanonical form according to the inverse Laplace transform formula \cite{Boissel1997}:

\begin{equation}
k_{diss}(E)=A^{diss}_{1000\mathrm{K}}\frac{\rho(E-E_a)}{\rho(E)},
\label{eq_kde}
\end{equation}
where $\rho(E)$ is the vibrational level density of 1-CNN$^+$, and we adopt the nominal value of the pre-exponential factor $A^{diss}_{1000\mathrm{K}}=k_B$(1000~K)$/h=2\times 10^{13}$~s$^{-1}$. The transition state temperatures $T^{\ddag}$ have been converted to vibrational energies $E$ according to the caloric curve computed directly from the vibrational frequencies \cite{Bull2019a} and including the finite heat bath correction \cite{Andersen2001} (see Supplemental Information).

We model the unimolecular dissociation rate in Fig.~\ref{fig_gamma} with the master equation approach described in Section~\ref{sec_model}. The calculated RF and IR cooling rate coefficients are plotted in Fig.~\ref{fig_rates}. Assuming the initial vibrational energy distribution is approximated by Boltzmann statistics, we find that an initial temperature of 1860(130)~K most closely reproduces the experimental dissociation rate. The simulated dissociation rate is given by the solid line labeled `Master' in Fig.~\ref{fig_gamma}. The simulated curve deviates from the experimental data for $t>20$~ms, which may be attributed to sequential fragmentation processes which have been observed in other ion-beam storage experiments with PAH ions \cite{Stockett2020b}.

The quenching of the dissociation rate is consistent with Recurrent Fluorescence from the lowest   electronic excited state $L_{\alpha}$.  To reproduce the measured dissociation rate, consideration of Herzberg-Teller vibronic coupling proved essential. This transition, like the first electronic transitions of most PAHs \cite{Hirata2003}, is forbidden by symmetry, but vibronic couplings relax this restriction. The calculated transition oscillator strength $f=0.011$ and hence RF rate coefficient (Eq.~\ref{eq_krf}) is two orders of magnitude lower if vibronic coupling is neglected (see Sec.~\ref{sec_model}). A simulation with this lower oscillator strength is shown by the dashed line labeled `Master (No HT)' in Fig.~\ref{fig_gamma} for the best-fitting initial temperature of 1310(50)~K. These findings are similar to those found in a previous study of perylene cations \cite{Stockett2020b}, which reported an RF rate consistent with an emission oscillator strength of 0.05(1), whereas most calculated values neglecting vibronic coupling are on the order of $10^{-4}$. The vibronic enhancement of $f$ in 1-CNN$^+$ is likely commonplace for a broad range of PAH molecules and decisive for the stabilization of such molecules in cold interstellar environments -- in stark contrast to what has been assumed previously.


The model survival probability of vibrationally hot 1-CNN$^+$ is plotted as a function of vibrational energy $E$ in the lower panel of Fig.~\ref{fig_rates}. The data points are determined from master equation simulations initialized to $\delta$-function vibrational energy distributions. Also included are simulations in which the RF rate coefficient is artificially set to zero, leaving only IR cooling. The solid line is the branching ratio $k_{rad}/(k_{diss}+k_{rad})$, where $k_{rad}=k_{RF}+k_{IR}$ or $k_{IR}$. The vertical step indicates the dissociation threshold ($k_{rad}=0$). The branching ratio when $k_{rad}=k_{RF}+k_{IR}$ agrees well with the full simulation, implying that a single RF photon stabilizes 1-CNN$^+$ \cite{Iida2021}. In contrast, when $k_{rad}=k_{IR}$, the branching ratio overestimates the survival probability from the simulations, as multiple IR photons are required to quench the decay. We conclude that while both cooling modes stabilize 1-CNN$^+$ ions with internal energies well above the dissociation threshold, RF does so much more efficiently for a critical range of internal excitation energies, as will be discussed below.

\section{Discussion}
\label{sec_disc}

While it has long been known that Herzberg-Teller vibronic coupling influences the optical spectra of PAHs \cite{Sklar1937}, the implications are not widely incorporated in astronomical contexts. In the present case, the large enhancement of the oscillator strength of the symmetry-forbidden $L_{\alpha}\leftarrow S_0$ transition leads to efficient radiative stabilization of 1-CNN$^+$ by recurrent fluorescence. This effect is generally important for PAHs \cite{Small1971,Negri1996}, and challenges the long-held assumption that small PAHs are rapidly depleted from the interstellar medium (ISM) by UV photodestruction \cite{Rapacioli2006,Montillaud2013}. Comparisons of measured and simulated optical spectra of PAHs to astronomical observations such as the Diffuse Interstellar Bands often discount the importance of the lowest energy transitions, on the basis of their low oscillator strengths in the Franck-Condon limit. Calculations including Herzberg-Teller coupling, benchmarked to quantitative laboratory data such as those presented here, will be crucial for such comparisons.

In the specific case of CNN in TMC-1, the present results suggest that the abundance predicted by astrochemical models may be too low due to both unrealistically low formation rates and unrealistically high destruction rates in the model \cite{McGuire2021}. McGuire \textit{et al.} acknowledge that small PAHs such as naphthalene may be inherited from the diffuse ISM at early stages of the evolution of TMC-1, but that formation of 1-CNN from these inherited PAHs is disfavored due to the inefficiency of radiative stabilization for small PAHs through IR-emission. Here, however, we have demonstrated that Recurrent Fluorescence provides efficient radiative stabilization of 1-CNN$^+$ with up to $\sim$5~eV of internal energy. Given the 8.6~eV ionization energy of 1-CNN \cite{Klasinc1983}, neutral 1-CNN should be resilient against dissociative ionization by 13.6~eV recombination radiation, which dominates the radiation field of molecular clouds \cite{Oberg2016}. RF is known to be an important mode of radiative cooling for the small unsubstituted PAH cations naphthalene \cite{Saito2020}, anthracene \cite{Martin2013}, and perylene \cite{Stockett2020b}, and is almost certainly more important for PAHs in general than predicted by calculations neglecting Herzberg-Teller coupling. The inherited abundances of small PAHs required to explain the observed amounts of CNNs in TMC-1 may thus not be as ``unrealistically large'' as previously thought \cite{McGuire2021}.

The astrochemical model employed by McGuire \textit{et al.} likely overestimates the rate of destruction of 1-CNN. In the model, naphthalene and CNNs are primarily destroyed in reactions with C$^+$, H$^+$, He$^+$, H$_3^+$, and H$_3$O$^+$ leading to small linear fragments such as \ce{C2H4}. However, some of the reactions included can be expected to predominantly lead to charge transfer rather than fragmentation \cite{Bohme1992,Petrie1993}, \textit{e.g.}

\begin{equation}
\ce{C+ + C10H7CN -> C + C10H7CN+}
\end{equation}
For this example, the excess energy of the reaction, \textit{i.e.} the 3.2~eV difference in ionization potentials, is insufficient to dissociate 1-CNN$^+$ at interstellar temperatures. The analogous reaction with H$^+$ (5.5~eV excess energy) should primarily lead, according to the present study, to HCN-loss (Eq.~\ref{eq_hcn}) and RF-stabilized 1-CNN$^+$, rather than disintegration into linear fragments. While 1-CNN$^+$ might be destroyed in dissociative recombination with electrons, radiative stabilization of the neutralized 1-CNN is possible \cite{Lacinbala2022}. Another process could involve 1-CNN formation by mutual neutralization (MN) of 1-CNN$^+$ with PAH anions, which have been suggested to be the dominant negative charge carriers in dark clouds \cite{Lepp1988}. MN is also possible with linear hydrocarbon anions like \ce{C6H-}, which have been identified in TMC-1 \cite{McCarthy2006} and have high electron binding energies \cite{McCarthy2006}, reducing the excess energy of the reaction. Finally, the proton-transfer reactions included in the model of McGuire \textit{et al.}, \textit{e.g.} with H$_3^+$, might lead to protonated PAHs, which are known to be stable and non-reactive \cite{Snow1998}, or to dehydrogenation, rather than carbon backbone fragmentation. 

In summary, the present laboratory study suggests that the situation in TMC-1 is closer to the ``best case scenario'' discussed by McGuire \textit{et al.}, with high inherited naphthalene abundance and no destruction by ions, than to their baseline assumptions of low inherited naphthalene and ion interactions leading always to linear fragments \cite{McGuire2021}. Additional laboratory studies of ion-neutral, electron-ion, and ion-ion reactions involving CNN under astrophysically relevant conditions should be undertaken to further constrain astrochemical models.




From the groundbreaking study of McGuire \textit{et al.} we now know from direct spectroscopic observations that a specific small PAH, 1-CNN, is present in the dark interstellar cloud TMC-1. In the present study we have shown that Recurrent Flouresence is sufficiently fast (due to Hertzberg-Teller vibronic coupling) to stabilize ionized 1-CNN molecules and that they may survive in harsh astrophysical environments, making their large presence in TMC-1 easier to understand. As similarly fast radiative cooling processes occur in a range of small PAHs \cite{Martin2013,Stockett2020b,Saito2020} it appears that the long-held assumption that small PAHs cannot survive in space has to be reconsidered.

\section{Methods}
\subsection{Experiments}
\label{sec_exp}
\begin{figure}
\includegraphics[width=\columnwidth]{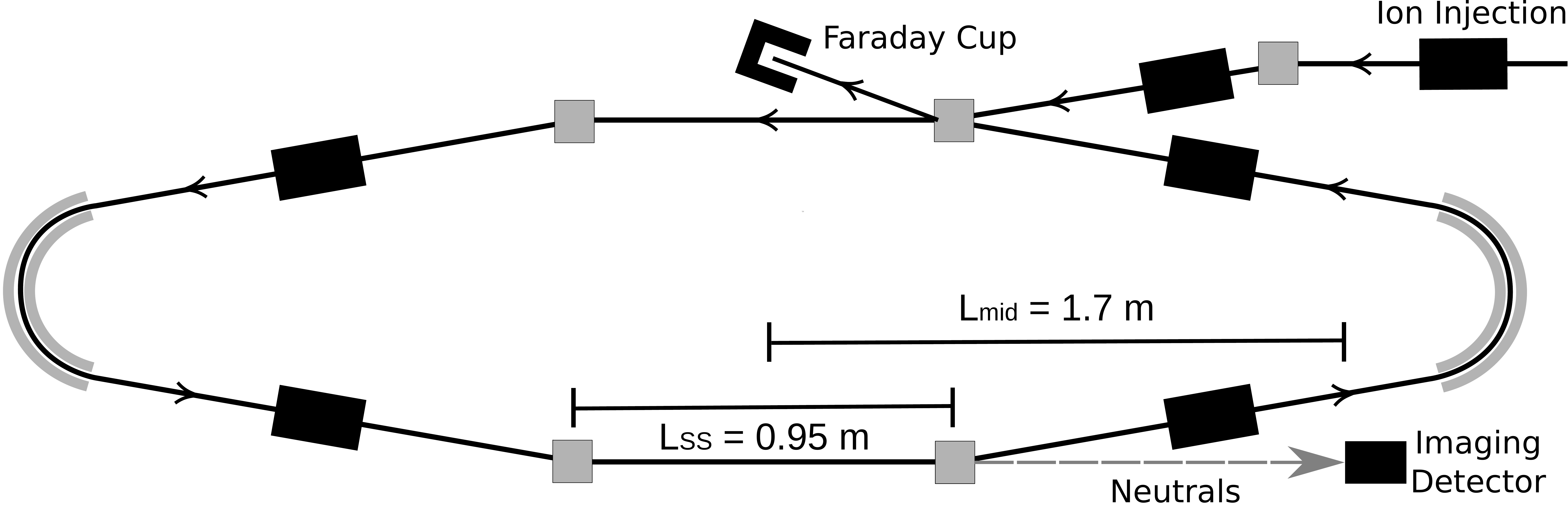}
\caption{\textbf{Schematic of the DESIREE electrostatic ion storage ring.} Beams of 1-CNN$^+$ with kinetic energies of 34~keV circulate along the trajectory indicated by the solid lines with arrows. Neutral products formed in the lower straight section follow the dashed line and are counted with the Imaging Detector.}
\label{fig_desiree}
\end{figure}

Experiments were conducted at the DESIREE (Double ElectroStatic Ion Ring ExpEriment) infrastructure at Stockholm University \cite{Thomas2011}. Cryogenic cooling of the DESIREE storage ring, which is schematically shown in Fig.~\ref{fig_desiree}, to $\approx$13~K results in a residual gas density on the order of $\sim 10^4$ cm$^{-3}$, consisting mostly of H$_2$ \cite{Schmidt2013}.

1-CNN (Sigma-Aldrich, $>96\%$) was sublimed from powder in a resistively heated oven coupled to an electron cyclotron resonance (ECR) ion source (Pantechnik Monogan) using helium as a support gas. Cations extracted from the source were accelerated to 34~keV kinetic energy. Mass-selected beams of cationic 1-CNN$^+$ ($m/z=153$) were stored in the DESIREE ion storage ring illustrated in Fig.~\ref{fig_desiree}.

After ion injection into DESIREE, neutral fragments are emitted from ions which retain sufficient internal energy from their formation process in the ion source. Neutrals formed in the observation arm (lower straight section in Fig.~\ref{fig_desiree}) of the storage ring continue with high velocity towards the position-sensitive Imaging Detector \cite{Eklund2020}, which utilizes custom ultra-high dynamic range micro-channel plates (MCPs, Photonis) that are suitable for high count rates at cryogenic temperatures. Electrons produced when neutral fragments strike the MCP are converted to optical photons with a phosphor screen. The resulting images are recorded through a vacuum window using a CMOS camera (Photon Focus MV1-D2048). 

Two types of neutral particle imaging experiment were performed. Firstly, a measurement was conducted with a continuous beam of 1-CNN$^+$ ions making a single pass around the ring. Including the transit time from the ion source to the storage ring, the ions in this experiment are decaying 120--124~$\mu$s after formation. For this single-pass measurement, a 0.5~mm aperture was inserted before the straight section of the storage ring to reduce the smearing of the neutral distributions due to the spatial extent of the beam. Secondly, measurements were performed using a stored beam circulating for 200~ms; no apertures were used. Exposures of $\Delta t=1.1$~ms duration were triggered every 2.5~ms. The imaging data were time-stamped and synchronized with the storage cycle, allowing for analysis of the time-dependence of the neutral particle distribution. Three-dimensional Newton spheres were reconstructed by applying an inverse Abel transform, using the `three-point' algorithm implemented in the PyAbel package \cite{pyabel}. The density distribution is related to the KER distribution by:

\begin{equation}
\epsilon(r_{3D}) = \frac{m_{neut}}{m_{cat}}E_{Acc}\left(\frac{r_{3D}}{L}\right)^2
\label{eq_epsilon}
\end{equation}
where $\epsilon(r_{3D})$ is the KER associated with a radial slice of the Newton sphere of radius $r_{3D}$, $m_{neut}$ and $m_{cat}$ are the masses of the neutral and cationic reaction products, $E_{Acc}=34$~keV is the beam energy, and $L$ is the distance traveled by the products from the point of reaction to the detector. For clarity of presentation, the KER distributions in Fig.~\ref{fig_ker} are plotted against an $\epsilon$ scale calculated according to Eq.~\ref{eq_epsilon} with $L=L_{mid}$, where $L_{mid}=1.7$~m is the distance from the detector to the mid-point of the observation arm. Our analysis accounts for dissociation occurring along the full length of the observation arm by summing contributions to the Newton sphere density distribution in the detector plane from points at distances in the range $L_{mid}\pm L_{SS}/2$, where $L_{SS}=0.95$~m is the length of the straight section seen by the detector (see Fig.~\ref{fig_desiree}). In the present case, the procedure gives a nearly insignificant correction relative to assuming all decays occur at $L_{mid}$.


The analysis of the dissociation rates and KER distributions considered only HCN-loss (Eq.~\ref{eq_hcn}). At the energies relevant to this study, H-loss is the only other competitive dissociation pathway, with \ce{C2H2}-loss being a minor channel \cite{West2019}. Low-mass H atoms emitted with kinetic energies around 200~eV are detected very inefficiently by the MCP detectors \cite{Stockett2020b} and are thus not expected to significantly contribute to the observed fragmentation rate or KER distributions.

\subsection{Kinetic Model}
\label{sec_model}
The quantitative kinetic model for PAH dissociation and radiative cooling used in this study has been detailed previously \cite{Stockett2019,Stockett2019a,Stockett2020b}. The dissociation rate coefficient $k_{diss}(E)$ is modeled using Eq.~\ref{eq_kde}. The vibrational level density $\rho(E)$ is computed using the Beyer-Swinehart algorithm \cite{Beyer1973}. Vibrational frequencies are calculated at the B3LYP/6-31G(d,p) level of Density Functional Theory (DFT) as implemented in Gaussian 16 \cite{g16}. 

The infrared radiative (vibrational) cooling rate coefficient $k_{IR}$ is calculated within the Simple Harmonic Cascade approximation \cite{Chandrasekaran2014}:

\begin{equation}
k_{IR}(E)=\sum_sk_s=\sum_sA_s^{IR}\sum_{v=1}^{v\leq E/h\nu_s} \frac{\rho(E-vh\nu_s)}{\rho(E)},
\label{eq_kir}
\end{equation}
where $v$ is the vibrational quantum number, and $h\nu_s$ and $A_s$ are the transition energy and Einstein coefficient of vibrational mode $s$. Previous studies have shown the infrared cooling rates predicted by this model reproduce eperimental data to within a factor of two \cite{Bull2019a,Stockett2020b}.   

The rate coefficient for RF is calculated using the expression \cite{Boissel1997}:

\begin{equation}
k_{RF}(E)=A^{RF}\frac{\rho(E-h\nu_{el})}{\rho(E)},
\label{eq_krf}
\end{equation}
where the electronic transition energies $h\nu_{el}$ and Einstein coefficients 
\begin{equation}
A^{RF}=\frac{2\pi\nu_{el}^2e^2}{\epsilon_0m_ec^3}f,
\end{equation}
where $f$ is the oscillator strength, are taken from density functional theory. Because PAHs are prototype examples of Herzberg-Teller activity \cite{Small1971,Bull2019} -- \textit{i.e.} coupled electronic and nuclear motion -- the oscillator strength for RF ($L_{\alpha}$ band fluorescence) was modeled using a Franck-Condon-Herzberg-Teller simulation \cite{Santoro2008} at the $\omega$B97X-D/cc-pVDZ level of theory to be $f=0.011$. Here, higher level EOM-CCSD/cc-pVDZ calculations of the excitation energy $h\nu_{el}$ were performed in CFOUR \cite{cfour}. The transition energy for RF is calculated at the equilibrium geometry of the lowest-lying $L_{\alpha}$ excited state and is $h\nu_{el}=1.10$ eV.  





The vibrational energy distribution, initially normalized such that $\int g(E,t=0)dE=1$, was propagated according to the Master Equation:

\begin{multline}
\frac{d}{dt}g(E,t)=-k_{diss}(E)g(E,t) \\ +\sum_{s} \left[ k_{s}(E+h\nu_s)g(E+h\nu_s,t)-k_{s}(E)g(E,t)\right] \\ + k_{RF}(E+h\nu_{el})g(E+h\nu_{el},t)-k_{RF}(E)g(E,t).
\label{eqn_master}
\end{multline}
The first term gives the depletion of the population by unimolecular dissociation. The first term in brackets represents $v+1\rightarrow v$ vibrational emission from levels above $E$ while the second is $v\rightarrow v-1$ emission to levels below $E$. The final two terms account for RF. The time step $dt$ is chosen to match the experimental data, with 32 extra points prior to the first experimental time bin to allow for the ion transit time from the ion source to the storage ring. The simulated dissociation rate is given by $\Gamma(t)=\int k_{diss}(E)g(E,t)dE$.

\section{Acknowledgements}

This work was supported by the Swedish Research Council (grant numbers 2016-03675, 2020-03437), the Knut and Alice Wallenberg Foundation (Grant No. 2018.0028), the Olle Engkvist Foundation (grant number 200-575), and the Swedish Foundation for International Collaboration in Research and Higher Education (STINT, grant number PT2017-7328 awarded to JNB and MHS). We acknowledge the DESIREE infrastructure for provisioning of facilities and experimental support, and thank the operators and technical staff for their invaluable assistance. The DESIREE infrastructure receives funding from the Swedish Research Council under the grant numbers 2017-00621 and 2021-00155. This article is based upon work from COST Action CA18212 - Molecular Dynamics in the GAS phase (MD-GAS), supported by COST (European Cooperation in Science and Technology).

\section{Data Availability}
The data that support the findings of this study are openly available in Zenodo at http://doi.org/10.5281/zenodo.XXX.


\begin{thebibliography}{10}
\expandafter\ifx\csname url\endcsname\relax
  \def\url#1{\texttt{#1}}\fi
\expandafter\ifx\csname urlprefix\endcsname\relax\def\urlprefix{URL }\fi
\providecommand{\bibinfo}[2]{#2}
\providecommand{\eprint}[2][]{\url{#2}}

\bibitem{Li2020}
\bibinfo{author}{Li, A.}
\newblock \bibinfo{title}{Spitzer’s perspective of polycyclic aromatic
  hydrocarbons in galaxies}.
\newblock \emph{\bibinfo{journal}{Nat. Astron.}} \textbf{\bibinfo{volume}{4}},
  \bibinfo{pages}{339--351} (\bibinfo{year}{2020}).

\bibitem{McGuire2021}
\bibinfo{author}{McGuire, B.~A.} \emph{et~al.}
\newblock \bibinfo{title}{Detection of two interstellar polycyclic aromatic
  hydrocarbons via spectral matched filtering}.
\newblock \emph{\bibinfo{journal}{Science}} \textbf{\bibinfo{volume}{371}},
  \bibinfo{pages}{1265--1269} (\bibinfo{year}{2021}).

\bibitem{McGuire2018}
\bibinfo{author}{McGuire, B.~A.} \emph{et~al.}
\newblock \bibinfo{title}{Detection of the aromatic molecule benzonitrile
  (c-c$_{6}$h$_{5}$cn) in the interstellar medium}.
\newblock \emph{\bibinfo{journal}{Science}} \textbf{\bibinfo{volume}{359}},
  \bibinfo{pages}{202--205} (\bibinfo{year}{2018}).

\bibitem{Rap2022}
\bibinfo{author}{Rap, D.~B.}, \bibinfo{author}{Schrauwen, J. G.~M.},
  \bibinfo{author}{Marimuthu, A.~N.}, \bibinfo{author}{Redlich, B.} \&
  \bibinfo{author}{Br{\"u}nken, S.}
\newblock \bibinfo{title}{Low-temperature nitrogen-bearing polycyclic aromatic
  hydrocarbon formation routes validated by infrared spectroscopy}.
\newblock \emph{\bibinfo{journal}{Nat. Astron.}}  (\bibinfo{year}{2022}).

\bibitem{Parker2012}
\bibinfo{author}{Parker, D. S.~N.} \emph{et~al.}
\newblock \bibinfo{title}{Low temperature formation of naphthalene and its role
  in the synthesis of {PAH}s ({P}olycyclic {A}romatic {H}ydrocarbons) in the
  interstellar medium}.
\newblock \emph{\bibinfo{journal}{Proc. Natl. Acad. Sci.}}
  \textbf{\bibinfo{volume}{109}}, \bibinfo{pages}{53--58}
  (\bibinfo{year}{2012}).

\bibitem{Burkhardt2021}
\bibinfo{author}{Burkhardt, A.~M.} \emph{et~al.}
\newblock \bibinfo{title}{Discovery of the pure {P}olycyclic {A}romatic
  {H}ydrocarbon indene (c-{C}$_9${H}$_8$) with {GOTHAM} observations of
  {TMC}-1}.
\newblock \emph{\bibinfo{journal}{Astrophys. J. Lett.}}
  \textbf{\bibinfo{volume}{913}}, \bibinfo{pages}{L18} (\bibinfo{year}{2021}).

\bibitem{Doddipatla2021}
\bibinfo{author}{Doddipatla, S.} \emph{et~al.}
\newblock \bibinfo{title}{Low-temperature gas-phase formation of indene in the
  interstellar medium}.
\newblock \emph{\bibinfo{journal}{Sci. Adv.}} \textbf{\bibinfo{volume}{7}}
  (\bibinfo{year}{2021}).

\bibitem{Lemmens2020}
\bibinfo{author}{Lemmens, A.~K.}, \bibinfo{author}{Rap, D.~B.},
  \bibinfo{author}{Thunnissen, J. M.~M.}, \bibinfo{author}{Willemsen, B.} \&
  \bibinfo{author}{Rijs, A.~M.}
\newblock \bibinfo{title}{Polycyclic {A}romatic {H}ydrocarbon formation
  chemistry in a plasma jet revealed by {IR-UV} action spectroscopy}.
\newblock \emph{\bibinfo{journal}{Nat. Commun.}} \textbf{\bibinfo{volume}{11}},
  \bibinfo{pages}{269} (\bibinfo{year}{2020}).

\bibitem{Rapacioli2006}
\bibinfo{author}{Rapacioli, M.} \emph{et~al.}
\newblock \bibinfo{title}{Formation and destruction of {P}olycyclic {A}romatic
  {H}ydrocarbon clusters in the interstellar medium}.
\newblock \emph{\bibinfo{journal}{Astron. Astrophys.}}
  \textbf{\bibinfo{volume}{460}}, \bibinfo{pages}{519--531}
  (\bibinfo{year}{2006}).

\bibitem{Montillaud2013}
\bibinfo{author}{Montillaud, J.}, \bibinfo{author}{Joblin, C.} \&
  \bibinfo{author}{Toublanc, D.}
\newblock \bibinfo{title}{Evolution of {P}olycyclic {A}romatic {H}ydrocarbons
  in photodissociation regions}.
\newblock \emph{\bibinfo{journal}{Astron. Astrophys.}}
  \textbf{\bibinfo{volume}{552}}, \bibinfo{pages}{A15} (\bibinfo{year}{2013}).

\bibitem{Martin2013}
\bibinfo{author}{Martin, S.} \emph{et~al.}
\newblock \bibinfo{title}{{Fast Radiative Cooling of Anthracene Observed in a
  Compact Electrostatic Storage Ring}}.
\newblock \emph{\bibinfo{journal}{Phys. Rev. Lett.}}
  \textbf{\bibinfo{volume}{110}}, \bibinfo{pages}{063003}
  (\bibinfo{year}{2013}).

\bibitem{Saito2020}
\bibinfo{author}{Saito, M.} \emph{et~al.}
\newblock \bibinfo{title}{Direct measurement of recurrent fluorescence emission
  from naphthalene ions}.
\newblock \emph{\bibinfo{journal}{Phys. Rev. A}}
  \textbf{\bibinfo{volume}{102}}, \bibinfo{pages}{012820}
  (\bibinfo{year}{2020}).

\bibitem{Nitzan1979}
\bibinfo{author}{Nitzan, A.} \& \bibinfo{author}{Jortner, J.}
\newblock \bibinfo{title}{Theory of inverse electronic relaxation}.
\newblock \emph{\bibinfo{journal}{J. Chem. Phys.}}
  \textbf{\bibinfo{volume}{71}}, \bibinfo{pages}{3524--3532}
  (\bibinfo{year}{1979}).

\bibitem{Ebara2016}
\bibinfo{author}{Ebara, Y.} \emph{et~al.}
\newblock \bibinfo{title}{Detection of recurrent fluorescence photons}.
\newblock \emph{\bibinfo{journal}{Phys. Rev. Lett.}}
  \textbf{\bibinfo{volume}{117}}, \bibinfo{pages}{133004}
  (\bibinfo{year}{2016}).

\bibitem{Ito2014}
\bibinfo{author}{Ito, G.} \emph{et~al.}
\newblock \bibinfo{title}{Cooling dynamics of photoexcited {C}$_6^-$ and
  {C}$_6${H}$^-$}.
\newblock \emph{\bibinfo{journal}{Phys. Rev. Lett.}}
  \textbf{\bibinfo{volume}{112}}, \bibinfo{pages}{183001}
  (\bibinfo{year}{2014}).

\bibitem{Chandrasekaran2014}
\bibinfo{author}{Chandrasekaran, V.} \emph{et~al.}
\newblock \bibinfo{title}{Determination of absolute recurrent fluorescence rate
  coefficients for {C}$_6^-$}.
\newblock \emph{\bibinfo{journal}{J. Phys. Chem. Lett.}}
  \textbf{\bibinfo{volume}{5}}, \bibinfo{pages}{4078--4082}
  (\bibinfo{year}{2014}).

\bibitem{West2019}
\bibinfo{author}{West, B.~J.}, \bibinfo{author}{Lesniak, L.} \&
  \bibinfo{author}{Mayer, P.~M.}
\newblock \bibinfo{title}{Why do large ionized {P}olycyclic {A}romatic
  {H}ydrocarbons not lose {C}$_2${H}$_2$?}
\newblock \emph{\bibinfo{journal}{J. Phys. Chem. A}}
  \textbf{\bibinfo{volume}{123}}, \bibinfo{pages}{3569--3574}
  (\bibinfo{year}{2019}).

\bibitem{Thomas2011}
\bibinfo{author}{Thomas, R.~D.} \emph{et~al.}
\newblock \bibinfo{title}{{The {D}ouble {E}lectro{S}tatic {I}on {R}ing
  {E}xp{E}riment: A unique cryogenic electrostatic storage ring for merged
  ion-beams studies}}.
\newblock \emph{\bibinfo{journal}{Rev. Sci. Instrum.}}
  \textbf{\bibinfo{volume}{82}}, \bibinfo{pages}{065112}
  (\bibinfo{year}{2011}).

\bibitem{Schmidt2013}
\bibinfo{author}{Schmidt, H.~T.} \emph{et~al.}
\newblock \bibinfo{title}{{First storage of ion beams in the Double
  Electrostatic Ion-Ring Experiment: DESIREE}}.
\newblock \emph{\bibinfo{journal}{Rev. Sci. Instrum.}}
  \textbf{\bibinfo{volume}{84}}, \bibinfo{pages}{055115}
  (\bibinfo{year}{2013}).

\bibitem{Hansen2001}
\bibinfo{author}{Hansen, K.} \emph{et~al.}
\newblock \bibinfo{title}{Observation of a 1/t decay law for hot clusters and
  molecules in a storage ring}.
\newblock \emph{\bibinfo{journal}{Phys. Rev. Lett.}}
  \textbf{\bibinfo{volume}{87}}, \bibinfo{pages}{123401}
  (\bibinfo{year}{2001}).

\bibitem{Andersen2001}
\bibinfo{author}{Andersen, J.}, \bibinfo{author}{Bonderup, E.} \&
  \bibinfo{author}{Hansen, K.}
\newblock \bibinfo{title}{On the concept of temperature for a small isolated
  system}.
\newblock \emph{\bibinfo{journal}{J. Chem. Phys.}}
  \textbf{\bibinfo{volume}{114}}, \bibinfo{pages}{6518--6525}
  (\bibinfo{year}{2001}).

\bibitem{Stockett2020b}
\bibinfo{author}{Stockett, M.~H.} \emph{et~al.}
\newblock \bibinfo{title}{Unimolecular fragmentation and radiative cooling of
  isolated {PAH} ions: A quantitative study}.
\newblock \emph{\bibinfo{journal}{J. Chem. Phys.}}
  \textbf{\bibinfo{volume}{153}}, \bibinfo{pages}{154303}
  (\bibinfo{year}{2020}).

\bibitem{Hansen2020}
\bibinfo{author}{Hansen, K.}
\newblock \bibinfo{title}{C$_{60}^-$ thermal electron-emission rate}.
\newblock \emph{\bibinfo{journal}{Phys. Rev. A}}
  \textbf{\bibinfo{volume}{102}}, \bibinfo{pages}{052823}
  (\bibinfo{year}{2020}).

\bibitem{Leyh1999}
\bibinfo{author}{Leyh, B.}
\newblock \bibinfo{title}{Ion dissociation kinetics in mass spectrometry}.
\newblock In \bibinfo{editor}{Lindon, J.~C.} (ed.)
  \emph{\bibinfo{booktitle}{Encyclopedia of Spectroscopy and Spectrometry
  (Second Edition)}}, \bibinfo{pages}{1127--1134} (\bibinfo{publisher}{Academic
  Press}, \bibinfo{address}{Oxford}, \bibinfo{year}{1999}),
  \bibinfo{edition}{second} edn.

\bibitem{Hansen2018}
\bibinfo{author}{Hansen, K.}
\newblock \bibinfo{title}{Tunneling and reflection in unimolecular reaction
  kinetic energy release distributions}.
\newblock \emph{\bibinfo{journal}{Chem. Phys. Lett.}}
  \textbf{\bibinfo{volume}{693}}, \bibinfo{pages}{66--71}
  (\bibinfo{year}{2018}).

\bibitem{So1988}
\bibinfo{author}{So, H.~Y.} \& \bibinfo{author}{Dunbar, R.~C.}
\newblock \bibinfo{title}{Time-resolved slow dissociation of benzonitrile ions
  by trapped-ion ion cyclotron resonance photodissociation}.
\newblock \emph{\bibinfo{journal}{J. Am. Chem. Soc.}}
  \textbf{\bibinfo{volume}{110}}, \bibinfo{pages}{3080--3083}
  (\bibinfo{year}{1988}).

\bibitem{Boissel1997}
\bibinfo{author}{Boissel, P.}, \bibinfo{author}{de~Parseval, P.},
  \bibinfo{author}{Marty, P.} \& \bibinfo{author}{Lefèvre, G.}
\newblock \bibinfo{title}{Fragmentation of isolated ions by multiple photon
  absorption: A quantitative study}.
\newblock \emph{\bibinfo{journal}{J. Chem. Phys.}}
  \textbf{\bibinfo{volume}{106}}, \bibinfo{pages}{4973--4984}
  (\bibinfo{year}{1997}).

\bibitem{Bull2019a}
\bibinfo{author}{Bull, J.~N.} \emph{et~al.}
\newblock \bibinfo{title}{Ultraslow radiative cooling of {C}$_n^-$ ($n =$
  3--5)}.
\newblock \emph{\bibinfo{journal}{J. Chem. Phys.}}
  \textbf{\bibinfo{volume}{151}}, \bibinfo{pages}{114304}
  (\bibinfo{year}{2019}).

\bibitem{Hirata2003}
\bibinfo{author}{Hirata, S.}, \bibinfo{author}{Head-Gordon, M.},
  \bibinfo{author}{Szczepanski, J.} \& \bibinfo{author}{Vala, M.}
\newblock \bibinfo{title}{Time-dependent density functional study of the
  electronic excited states of {P}olycyclic {A}romatic {H}ydrocarbon radical
  ions}.
\newblock \emph{\bibinfo{journal}{J. Phys. Chem. A}}
  \textbf{\bibinfo{volume}{107}}, \bibinfo{pages}{4940--4951}
  (\bibinfo{year}{2003}).

\bibitem{Iida2021}
\bibinfo{author}{Iida, S.} \emph{et~al.}
\newblock \bibinfo{title}{{IR}-photon quenching of delayed electron detachment
  from hot pentacene anions}.
\newblock \emph{\bibinfo{journal}{Phys. Rev. A}}
  \textbf{\bibinfo{volume}{104}}, \bibinfo{pages}{043114}
  (\bibinfo{year}{2021}).

\bibitem{Sklar1937}
\bibinfo{author}{Sklar, A.~L.}
\newblock \bibinfo{title}{Theory of color of organic compounds}.
\newblock \emph{\bibinfo{journal}{J. Chem. Phys.}}
  \textbf{\bibinfo{volume}{5}}, \bibinfo{pages}{669--681}
  (\bibinfo{year}{1937}).

\bibitem{Small1971}
\bibinfo{author}{Small, G.~J.}
\newblock \bibinfo{title}{Herzberg-{T}eller vibronic coupling and the
  {D}uschinsky effect}.
\newblock \emph{\bibinfo{journal}{J. Chem. Phys.}}
  \textbf{\bibinfo{volume}{54}}, \bibinfo{pages}{3300--3306}
  (\bibinfo{year}{1971}).

\bibitem{Negri1996}
\bibinfo{author}{Negri, F.} \& \bibinfo{author}{Zgierski, M.~Z.}
\newblock \bibinfo{title}{Vibronic structure of the emission spectra from
  single vibronic levels of the {S}$_1$ manifold in naphthalene: Theoretical
  simulation}.
\newblock \emph{\bibinfo{journal}{J. Chem. Phys.}}
  \textbf{\bibinfo{volume}{104}}, \bibinfo{pages}{3486--3500}
  (\bibinfo{year}{1996}).

\bibitem{Klasinc1983}
\bibinfo{author}{Klasinc, L.}, \bibinfo{author}{Kovac, B.} \&
  \bibinfo{author}{Gusten, H.}
\newblock \bibinfo{title}{Photoelectron spectra of acenes. {E}lectronic
  structure and substituent effects}.
\newblock \emph{\bibinfo{journal}{Pure Appl. Chem.}}
  \textbf{\bibinfo{volume}{55}}, \bibinfo{pages}{289--298}
  (\bibinfo{year}{1983}).

\bibitem{Oberg2016}
\bibinfo{author}{\"{O}berg, K.~I.}
\newblock \bibinfo{title}{Photochemistry and astrochemistry: Photochemical
  pathways to interstellar complex organic molecules}.
\newblock \emph{\bibinfo{journal}{Chem. Rev.}} \textbf{\bibinfo{volume}{116}},
  \bibinfo{pages}{9631--9663} (\bibinfo{year}{2016}).

\bibitem{Bohme1992}
\bibinfo{author}{Bohme, D.~K.}
\newblock \bibinfo{title}{{PAH} [{P}olycyclic {A}romatic {H}ydrocarbons] and
  fullerene ions and ion/molecule reactions in interstellar and circumstellar
  chemistry}.
\newblock \emph{\bibinfo{journal}{Chem. Rev.}} \textbf{\bibinfo{volume}{92}},
  \bibinfo{pages}{1487--1508} (\bibinfo{year}{1992}).

\bibitem{Petrie1993}
\bibinfo{author}{Petrie, S.}, \bibinfo{author}{Javahery, G.},
  \bibinfo{author}{Fox, A.} \& \bibinfo{author}{Bohme, D.~K.}
\newblock \bibinfo{title}{Novel charge-transfer electron-detachment reaction
  for the production of naphthalene dications at thermal energy}.
\newblock \emph{\bibinfo{journal}{J. Phys. Chem.}}
  \textbf{\bibinfo{volume}{97}}, \bibinfo{pages}{5607--5610}
  (\bibinfo{year}{1993}).

\bibitem{Lacinbala2022}
\bibinfo{author}{Lacinbala, O.} \emph{et~al.}
\newblock \bibinfo{title}{Radiative relaxation in isolated large carbon
  clusters: Vibrational emission versus recurrent fluorescence}.
\newblock \emph{\bibinfo{journal}{J. Chem Phys.}}
  \textbf{\bibinfo{volume}{156}}, \bibinfo{pages}{144305}
  (\bibinfo{year}{2022}).

\bibitem{Lepp1988}
\bibinfo{author}{Lepp, S.} \& \bibinfo{author}{Dalgarno, A.}
\newblock \bibinfo{title}{{P}olycyclic {A}romatic {H}ydrocarbons in
  interstellar chemistry}.
\newblock \emph{\bibinfo{journal}{Astrophys. J.}}
  \textbf{\bibinfo{volume}{324}}, \bibinfo{pages}{553} (\bibinfo{year}{1988}).

\bibitem{McCarthy2006}
\bibinfo{author}{McCarthy, M.~C.}, \bibinfo{author}{Gottlieb, C.~A.},
  \bibinfo{author}{Gupta, H.} \& \bibinfo{author}{Thaddeus, P.}
\newblock \bibinfo{title}{Laboratory and astronomical identification of the
  negative molecular ion {C}$_6${H}$^-$}.
\newblock \emph{\bibinfo{journal}{Astrophy. J. Lett.}}
  \textbf{\bibinfo{volume}{652}}, \bibinfo{pages}{L141} (\bibinfo{year}{2006}).

\bibitem{Snow1998}
\bibinfo{author}{Snow, T.~P.}, \bibinfo{author}{Page, V.~L.},
  \bibinfo{author}{Keheyan, Y.} \& \bibinfo{author}{Bierbaum, V.~M.}
\newblock \bibinfo{title}{The interstellar chemistry of {PAH} cations}.
\newblock \emph{\bibinfo{journal}{Nature}} \textbf{\bibinfo{volume}{391}},
  \bibinfo{pages}{259--260} (\bibinfo{year}{1998}).

\bibitem{Eklund2020}
\bibinfo{author}{Eklund, G.} \emph{et~al.}
\newblock \bibinfo{title}{Cryogenic merged-ion-beam experiments in {DESIREE}:
  Final-state-resolved mutual neutralization of {Li}$^{+}$ and
  {D}$^{\ensuremath{-}}$}.
\newblock \emph{\bibinfo{journal}{Phys. Rev. A}}
  \textbf{\bibinfo{volume}{102}}, \bibinfo{pages}{012823}
  (\bibinfo{year}{2020}).

\bibitem{pyabel}
\bibinfo{author}{Gibson, S.} \emph{et~al.}
\newblock \bibinfo{title}{Pyabel: v0.8.4} (\bibinfo{year}{2021}).

\bibitem{Stockett2019}
\bibinfo{author}{Stockett, M.~H.}, \bibinfo{author}{Bj{\"o}rkhage, M.},
  \bibinfo{author}{Cederquist, H.}, \bibinfo{author}{Schmidt, H.} \&
  \bibinfo{author}{Zettergren, H.}
\newblock \bibinfo{title}{Storage time dependent photodissociation action
  spectroscopy of {P}olycyclic {A}romatic {H}ydrocarbon cations in the
  cryogenic electrostatic storage ring {DESIREE}}.
\newblock \emph{\bibinfo{journal}{Faraday Discuss.}}
  \textbf{\bibinfo{volume}{217}}, \bibinfo{pages}{126--137}
  (\bibinfo{year}{2019}).

\bibitem{Stockett2019a}
\bibinfo{author}{Stockett, M.~H.}, \bibinfo{author}{Bj{\"o}rkhage, M.},
  \bibinfo{author}{Cederquist, H.}, \bibinfo{author}{Schmidt, H.~T.} \&
  \bibinfo{author}{Henning, Z.}
\newblock \bibinfo{title}{Intrinsic absorption profile and radiative cooling
  rate of a {PAH} cation revealed by action spectroscopy in the cryogenic
  electrostatic storage ring {DESIREE}}.
\newblock \emph{\bibinfo{journal}{Proc. Int. Astron. Union}}
  \textbf{\bibinfo{volume}{15}}, \bibinfo{pages}{127–131}
  (\bibinfo{year}{2019}).

\bibitem{Beyer1973}
\bibinfo{author}{Beyer, T.} \& \bibinfo{author}{Swinehart, D.~F.}
\newblock \bibinfo{title}{Algorithm 448: Number of multiply-restricted
  partitions}.
\newblock \emph{\bibinfo{journal}{Commun. ACM}} \textbf{\bibinfo{volume}{16}},
  \bibinfo{pages}{379} (\bibinfo{year}{1973}).

\bibitem{g16}
\bibinfo{author}{Frisch, M.~J.} \emph{et~al.}
\newblock \bibinfo{title}{Gaussian~16 {R}evision {B}.01}
  (\bibinfo{year}{2016}).
\newblock \bibinfo{note}{Gaussian Inc. Wallingford CT 2016}.

\bibitem{Bull2019}
\bibinfo{author}{Bull, J.~N.} \emph{et~al.}
\newblock \bibinfo{title}{Photodetachment and photoreactions of substituted
  naphthalene anions in a tandem ion mobility spectrometer}.
\newblock \emph{\bibinfo{journal}{Faraday Discuss.}}
  \textbf{\bibinfo{volume}{217}}, \bibinfo{pages}{34--46}
  (\bibinfo{year}{2019}).

\bibitem{Santoro2008}
\bibinfo{author}{Santoro, F.}, \bibinfo{author}{Lami, A.},
  \bibinfo{author}{Improta, R.}, \bibinfo{author}{Bloino, J.} \&
  \bibinfo{author}{Barone, V.}
\newblock \bibinfo{title}{Effective method for the computation of optical
  spectra of large molecules at finite temperature including the {D}uschinsky
  and {H}erzberg-{T}eller effect: The {Q}$_x$ band of porphyrin as a case
  study}.
\newblock \emph{\bibinfo{journal}{J. Chem. Phys.}}
  \textbf{\bibinfo{volume}{128}}, \bibinfo{pages}{224311}
  (\bibinfo{year}{2008}).

\bibitem{cfour}
\bibinfo{author}{Stanton, J.~F.} \emph{et~al.}
\newblock \bibinfo{title}{{CFOUR, Coupled-Cluster techniques for Computational
  Chemistry, a quantum-chemical program package}}.

\end{thebibliography}

\end{document}